\documentclass[a4paper, 12 pt]{article}
\usepackage[utf8]{inputenc}
\usepackage{jheppub}
\pdfoutput=1
\usepackage[english]{babel}
\usepackage{amsmath}
\usepackage{amssymb}
\usepackage{physics}
\usepackage{graphicx}
\usepackage{caption}
\usepackage{subcaption}
\usepackage{hyperref}

\title{\vspace{-2cm}
    \begin{flushright}
        {\normalsize INR-TH-2023-009}
    \end{flushright}
    \vspace{0.5cm} Dark photon production via elastic proton bremsstrahlung with non-zero momentum transfer}

\author[a,b]{D. Gorbunov,}
\author[a,c]{ E. Kriukova}

\affiliation[a]{Institute for Nuclear Research of the Russian Academy of Sciences, \\
60th October Anniversary pr-ct 7a, Moscow 117312, Russia}
\affiliation[b]{Landau Phystech School of Physics and Research, Moscow Institute of Physics and Technology, \\
Institutskiy per. 9, Dolgoprudny 141700, Russia}
\affiliation[c]{Faculty of Physics, Lomonosov Moscow State University,\\
Leninskiye Gory 1-2, Moscow 119991, Russia}

\emailAdd{gorby@inr.ac.ru}
\emailAdd{kryukova.ea15@physics.msu.ru}

\abstract{We explore hypothetical vector particles, dark photons $\gamma'$, which mix with the Standard Model photons and thus mediate interactions with charged particles into the hidden sector. We study the elastic proton bremsstrahlung of dark photons with masses 0.4--1.8 GeV, relevant for direct searches with proton accelerators. A key feature of our calculation is that it explicitly considers the non-zero momentum transfer between protons in the process  $pp\rightarrow pp\gamma'$. We compare the obtained differential and full bremsstrahlung cross sections with the results of other authors. Our calculation agrees well (up to 3--9\% corrections) with the Weizsacker-Williams approximation that confirms its applicability for proton beams. Then we refine predictions for the dark photon production with proton beams of energy 30\,GeV, 70\,GeV, 120\,GeV and 400\,GeV relevant for past, present and future experiments considered in literature.}

\arxivnumber{2306.15800}

\begin{document}
\maketitle
\flushbottom
\section{Introduction}
A variety of experimental data including observations of galaxy rotation curves \cite{Rubin:1970zza,deMartino:2020gfi} and neutrino oscillations \cite{Super-Kamiokande:1998kpq,Ahlers:2018mkf} indicate the need to extend the Standard Model (SM) of particle physics and explore yet unknown particles of dark sector singlet with respect to the SM gauge group. One common approach is to couple the dark sector to the SM via special new particles, mediators, that are able to interact with both of them \cite{Essig:2013lka}. There exist only three ways to write down the renormalizable interaction of the SM fields with a mediator: scalar, vector and fermion portals \cite{Lanfranchi:2020crw}. In this study we concentrate on the vector portal in which mediator is so-called dark photon.

We consider only the part of the model responsible for the portal-type interaction, i.e.  the SM Lagrangian $\mathcal{L}_{SM}$ is extended as follows \cite{Okun:1982xi,Galison:1983pa,Holdom:1985ag},
\begin{equation}
\label{eq:Lagr}
\mathcal{L}=\mathcal{L}_{SM}-\frac{1}{4}F'_{\mu \nu}F'^{\mu \nu}+\frac{\epsilon}{2}F'_{\mu\nu}B^{\mu \nu}+\frac{m_{\gamma'}^2}{2}A'_\mu A'^{\mu},
\end{equation}
where $F'_{\mu\nu}$ is the dark photon strength tensor, $B_{\mu\nu}$ is the SM hypercharge strength tensor, $\epsilon$ is the kinetic mixing parameter and $m_{\gamma'}$ is the dark photon mass. Due to mixing with the SM photon the dark photon couples to all SM particles, initially charged with $q$ under $U(1)_{em}$, with charge $\epsilon q$  \cite{Filippi:2020kii}. In what follows we discuss the dark photon production by accelerated protons incident on target. The corresponding center-of-mass energies are much lower than the electroweak energy scale, so the $Z$-boson mixing induced by the portal coupling\,\eqref{eq:Lagr} can be safely neglected.  

The vector portal has been intensively studied in experiments with electron beam dumps (NA64 \cite{NA64:2016oww,NA64:2017vtt,NA64:2018lsq,Banerjee:2019pds,NA64:2019auh}, E141 \cite{Riordan:1987aw,Bjorken:2009mm,Andreas:2012mt}), proton beam dumps (PS191 \cite{Bernardi:1985ny,Gninenko:2011uv}, NOMAD \cite{Gninenko:2011uv,NOMAD:2001eyx}, CHARM \cite{CHARM:1985nku,Gninenko:2012eq}), at $e^+e^-$ colliders (BaBar \cite{BaBar:2014zli,BaBar:2017tiz}, KLOE \cite{KLOE-2:2011hhj,KLOE-2:2012lii,Anastasi:2015qla,KLOE-2:2016ydq}) and at $pp$ colliders (LHCb \cite{LHCb:2017trq,LHCb:2019vmc}, CMS \cite{CMS:2018jid,CMS:2019ajt,CMS:2019buh,CMS:2020krr}),  see \cite{Essig:2013lka,Raggi:2015yfk,Agrawal:2021dbo} for the review on dark photon experimental searches. However for dark photon masses above 0.8 GeV experimental constraints are still quite weak, $\epsilon < 5\cdot 10^{-3}$ \cite{Graham:2021ggy}. Therefore several experiments under construction, such as DUNE \cite{DUNE:2020fgq,Breitbach:2021gvv}, T2K \cite{deNiverville:2016rqh} and proposals such as SHiP \cite{Gorbunov:2014wqa,SHiP:2015vad,SHiP:2020vbd} aim also at searching for dark photons of mass about 1\,GeV produced in $pp$ collisions \cite{Batell:2022dpx}. In order to estimate the sensitivity of these projects one needs to study the phenomenology of ${\cal O}(1)$\,GeV dark photon, in particular its production modes.

There are three mechanisms of dark photon production in proton-proton collisions \cite{Miller:2021ycl}. The dominant one is determined by the value of dark photon mass $m_{\gamma'}$. Dark photons lighter than about 0.4 GeV are mostly produced in meson decays $\pi^0/\eta\rightarrow \gamma' \gamma$ due to dark photon mixing with the SM photon. The QCD Drell-Yan process $q\bar{q}\rightarrow \gamma'$ is the most important mode for dark photons with mass $m_{\gamma'}\gtrsim 1.8$ GeV. In the intermediate mass range 0.4--1.8\,GeV the dark photon production is mainly due to proton bremsstrahlung $pp\rightarrow\gamma'pp$. In this study we consider dark photon production in elastic proton bremsstrahlung. Our main result is a new estimate for its cross section taking into account a realistic non-zero momentum transfer between the protons.

The paper is organized as follows. In section \ref{overview} we give a brief overview of the existing estimates of bremsstrahlung cross section. Section \ref{calculation} contains our calculation of the proton bremsstrahlung cross section with non-zero momentum transfer. In section \ref{comparison} we present the numerical results for typical dark photon masses and compare them with other approximations. Section \ref{discussion} summarizes our findings and future prospects.  

\section{Overview of the previously suggested methods} \label{overview}
One of the first successful attempts to calculate the electron bremsstrahlung is given in \cite{Kim:1973he}. There the modified Weizsacker-Williams approximation is formulated assuming the one-photon exchange in the process $e(a)p(P_i)\rightarrow \gamma'(b) e(c)p(P_f)$. The photon propagator gives the contribution $1/t^2$ to the square of matrix element, where $t\equiv-q^2$ and $q$ is the momentum of virtual photon. Then the main contribution comes from a phase space region near the minimal value of $t$
\begin{equation}
\sqrt{t_{\text{min}}}=\frac{(a\cdot b)-m^2_{\gamma '}/2}{a_0-b_0}.
\end{equation}
In the lab frame this happens when the difference of 3-dimensional vectors 
$\vec{a}-\vec{b}$ and vector $\vec{c}$ are parallel,  
$\vec{a}-\vec{b}\parallel\vec{c}$. The differential bremsstrahlung cross section is approximated then by the differential cross section of the subprocess $e(a)\gamma(q)\rightarrow \gamma'(b)e(c)$ taken at $t=t_{\text{min}}$
\begin{equation}
\left[ \frac{\dd^2 \sigma(ep\rightarrow \gamma 'ep)}{\dd (a\cdot b) \dd (b\cdot P_i)} \right]_{WW}= \left[\frac{\dd \sigma(e\gamma \rightarrow \gamma' e)}{\dd (a\cdot b)}\right]_{t=t_{\text{min}}}\frac{\alpha}{\pi}\frac{\chi}{(c \cdot P_i)},
\end{equation}
where $\alpha$ is the fine-structure constant, $\chi$ is the flux of photons emitted by the target proton 
\begin{equation}
\chi=\int\displaylimits_{t_\text{min}}^{t_\text{max}}\frac{t-t_\text{min}}{t^2}G_2(t)\, \dd t
\end{equation}
and $G_2(t)$ is the proton electric form factor depending on the energy transfer as presented in \cite{Kim:1973he}. This approximation was used to calculate the dark photon production cross section in electron bremsstrahlung \cite{Bjorken:2009mm}. The validity of Weizsacker-Williams approximation for dark photon production in electron bremsstrahlung was questioned in \cite{Liu:2017htz}.

Recently the Weizsacker-Williams approximation has been applied to proton bremsstrahlung $p(p)p(P_i)\rightarrow\gamma'(k)p(p')p(P_f)$ in the case of elastic proton scattering by Foroughi-Abari and Ritz \cite{Foroughi-Abari:2021zbm}. In this approach the protons exchange virtual effective vector bosons, the so-called Donnachie-Landshoff pomerons $\mathbb{P}$, which have effective propagators and proton-pomeron vertices obtained from fits to  experimental data on elastic proton scattering \cite{Donnachie:1992ny,Donnachie:2013xia}. The differential cross section of proton bremsstrahlung is given by
\begin{equation} \label{eq:Ritz}
\left[\frac{\dd^2 \sigma(pp\rightarrow \gamma' pp)}{\dd z \dd k^2_\perp}\right]_{WW}=\frac{\epsilon^2\alpha z(1-z)}{16\pi^2H^2} A_{2\rightarrow 2} \chi_\mathbb{P} \abs{F_{\text{VMD}}}^2 F_{\text{virt}}^2\,.
\end{equation}
Here $z\equiv k_z/p_z$ is the ratio of projections of dark photon 3-momentum and incident proton 3-momentum on the beam axis, $k_{\perp}$ is the transverse momentum of dark photon, 
\begin{equation} \label{eq:H}
H\equiv k^2_\perp+(1-z)m^2_{\gamma '}+z^2M^2,
\end{equation}
and $M$ is the proton mass. All particle momenta are assigned below in \eqref{eq:momenta}. 
In the limit $t\rightarrow 0$ the squared and averaged matrix element of the subprocess $p(p)\mathbb{P}(q)\rightarrow\gamma'(k)p(p')$ reads
\begin{equation} \label{eq:A22}
A_{2\rightarrow 2}=2\left(\frac{1+\left(1-z\right)^2}{1-z} - 2\left(2M^2+m^2_{\gamma'}\right)\frac{k^2_\perp z^2}{H^2}\right).
\end{equation} 
The pomeron flux from the target proton is
\begin{equation} \label{eq:chi}
\chi_{\mathbb{P}}=\int\displaylimits_{t_\text{min}}^{t_\text{max}} \left(t-t_{\text{min}}\right) \abs{\mathcal{M}_{\mathbb{P}}}^2\dd t,
\end{equation}
where $\mathcal{M}_{\mathbb{P}}$ is the sum of matrix elements describing single pomeron, double pomeron and triple gluon exchanges. The latter contribution can be neglected for the interesting here case of $|t|\lesssim 1$\,GeV$^2$. The cross section \eqref{eq:Ritz} includes two new form factors: the form factor accounting for possible dark photon mixing with $\rho$- and $\omega$-family mesons according to the vector meson dominance (VMD) hypothesis \cite{Faessler:2009tn},
\begin{equation} \label{eq:vmdff}
F_{\text{VMD}}\equiv\sum_{i}\frac{f_i m_i^2}{m_i^2-m_{\gamma '}^2-im_i\Gamma_i},
\end{equation}
with the vector meson fit parameters $f_i$, masses $m_i$ and decay widths $\Gamma_i$ given in table~\ref{tab:VMD}, 
\begin{table}[]
    \begin{center}
    \begin{tabular}{|c|c|c|c|c|c|c|}
    \hline
    & $\rho$ & $\omega$ & $\rho'$ & $\omega'$ & $\rho''$ & $\omega''$ \\
    \hline
    $f_i$ & 0.616 & 1.011 & 0.223 & -0.881 & -0.339 & 0.369 \\
    \hline
    $m_i$, GeV & \multicolumn{2}{|c|}{0.770} & \multicolumn{2}{|c|}{1.25} & \multicolumn{2}{|c|}{1.45} \\
    \hline
    $\Gamma_i$, GeV & 0.150 & 0.0085 & \multicolumn{2}{|c|}{0.300} & \multicolumn{2}{|c|}{0.500} \\
    \hline
    \end{tabular}    
    \end{center}
    \caption{The parameters of $\rho$- and $\omega$-family mesons adopted in the VMD form factor \eqref{eq:vmdff}.}
    \label{tab:VMD}
\end{table}
and the off-shell hadronic form factor with the hard scale $\Lambda=1.5\text{ GeV}$ \cite{Feuster:1998cj},    
\begin{equation} \label{eq:virtff}
F_{\text{virt}}\equiv \frac{\Lambda^4}{\Lambda^4+\left((p-k)^2-M^2\right)^2},
\end{equation}
penalising for the proton compositeness.  

Another widely used approximation for the proton bremsstrahlung has been obtained by Blumlein and Brunner in \cite{Blumlein:2013cua}. Although in the original paper the authors study the inelastic bremsstrahlung with inclusive cross section, we find that the similar result can be obtained in the elastic case along the same lines. The idea of the calculation is the following. Proton-proton interaction is considered as the exchange of hypothetical massless vector particles $b$ (in full analogy with virtual photons in the electron bremsstrahlung \cite{Kim:1973he}). In contrast to pomerons in \cite{Foroughi-Abari:2021zbm}, these particles have usual massless vector propagator proportional to $1/q^2$. Firstly, using the Weizsacker-Williams approximation the bremsstrahlung matrix element is related to the amplitude of the $2\rightarrow 2$ process $pb\rightarrow \gamma'p$. Secondly, the probability of the subprocess $p\rightarrow\gamma'p$ is extracted from the $pb\rightarrow \gamma'p$ cross section 
and the hypothetical boson momentum $q^\mu$ is set exactly to 0 revealing 
\begin{equation} \label{eq:BB1}
\left[ \frac{\dd^2 \sigma (pp\rightarrow \gamma'pp)}{\dd z \dd k^2_\perp}\right]_{BB} = w_{p\gamma'}(z, k_\perp^2)\sigma_{pp}(\bar{s}),
\end{equation}
where the splitting function for $p\rightarrow\gamma'p$ is given by \cite{Blumlein:2013cua}
\begin{multline} \label{eq:BB2}
	w_{p\gamma'}(z, k_\perp^2)=\frac{\epsilon^2 \alpha}{2\pi H} \left[\frac{1+(1-z)^2}{z}-2z(1-z)\left(\frac{2M^2+m^2_{\gamma'}}{H}-z^2\frac{2M^4}{H^2}\right)+
	\right.\\\left. + 2z(1-z)(1+(1-z)^2)\frac{M^2 m_{\gamma'}^2}{H^2}+2z(1-z)^2\frac{m^4_{\gamma'}}{H^2}\right],
\end{multline}
$\sigma_{pp}(\bar{s})$ is the full elastic proton scattering cross section and $\sqrt{\bar{s}}$ is the center-of-mass energy of two protons which exchange the hypothetical boson. Comparing \eqref{eq:Ritz}-\eqref{eq:chi} with \eqref{eq:BB1}, \eqref{eq:BB2} one can see that these two results show completely different behavior as functions of proton and dark photon masses.

\begin{figure}[t]
	\begin{center}
		\includegraphics[width=0.5\textwidth]{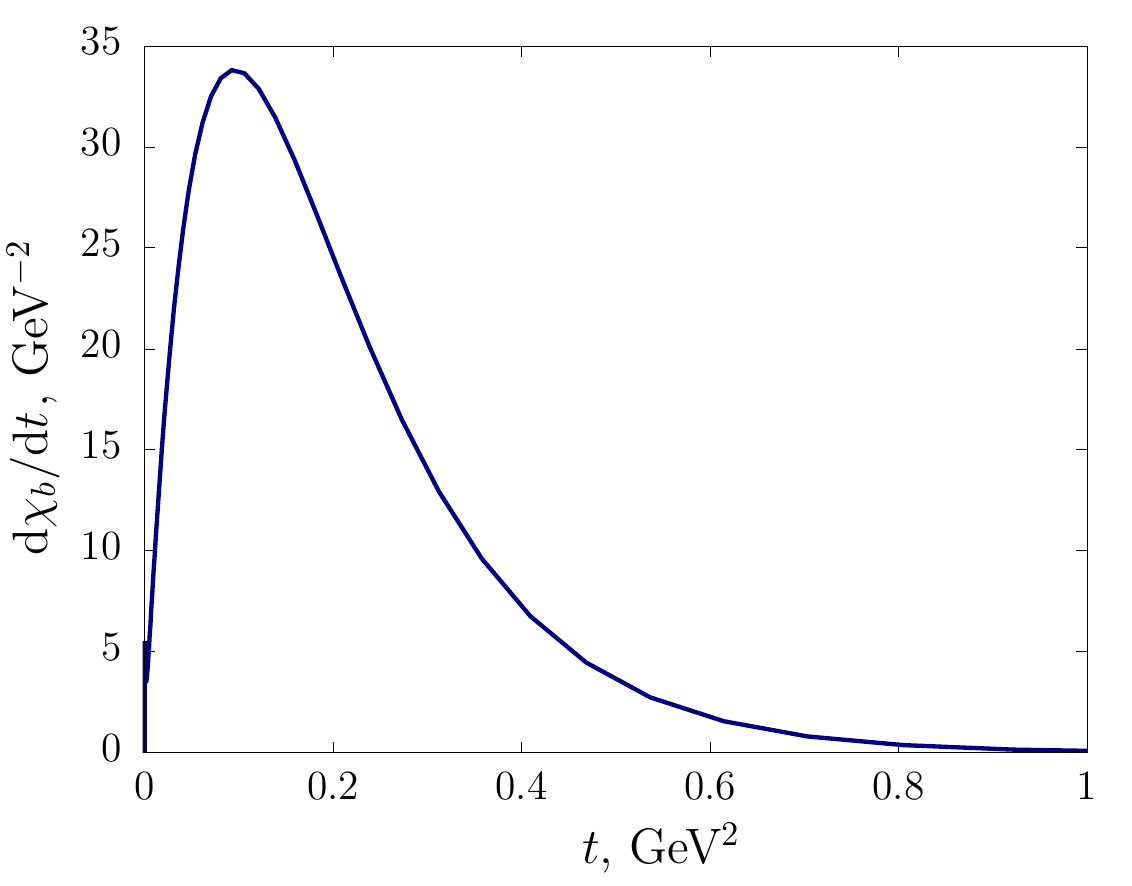}
		\caption{The differential flux of hypothetical bosons in proton bremsstrahlung \eqref{WW-boson-flux} according to the Weizsacker-Williams approximation for $P=120\text{ GeV}$, $m_{\gamma '}=1\text{ GeV}$, $z=0.5$, $k_\perp=0.1\text{ GeV}$.}
		\label{fig:chi}
	\end{center}
\end{figure}

All the described above approaches explicitly put either the momentum of hypothetical virtual boson or its square to zero assuming that the maximum of hypothetical boson flux is reached at very small $t$. Although this is indeed the case for electron bremsstrahlung, the proton bremsstrahlung requires more detailed treatment. On the base of the analysis performed in this paper, in figure~\ref{fig:chi}
we plot the differential flux of hypothetical bosons
\begin{equation}
\label{WW-boson-flux}
	\frac{\dd \chi_b}{\dd t}=\left(t-t_{\text{min}}\right)\abs{\mathcal{M}_{pp}}^2,
\end{equation}
where $\abs{\mathcal{M}_{pp}}^2\equiv 4\pi \dd \sigma_{pp}/\dd q^2$, the differential cross section for the elastic proton scattering $\dd \sigma_{pp}/\dd q^2$ is given by \eqref{eq:elasticamp} and we take the numerical fit for the amplitude of elastic scattering from \cite{ParticleDataGroup:2016lqr}. For the center-of-mass energy of two protons squared we take its typical value in the proton bremsstrahlung $\bar{s}=2M^2+2MP(1-z)-H/z$ with $P=120\text{ GeV}$, $m_{\gamma '}=1\text{ GeV}$, $z=0.5$ and $k_\perp=0.1\text{ GeV}$. It can be easily seen that the maximum of such hypothetical boson flux is approached not at zero, but at about QCD scale, $\sqrt{t}\simeq0.35\text{ GeV}$, which is numerically not a negligible scale in this problem. 
This observation suggests a proper refinement of the previous estimates of the dark photon production via the proton bremsstrahlung. 

Thus in the next section we perform the calculation of proton bremsstrahlung cross section considering non-zero momentum $q$ of the hypothetical virtual boson. Our main goal is to check whether the Weizsacker-Williams approximation still gives the reliable result in the case of incident proton beam.


\section{The bremsstrahlung cross section with non-zero momentum transfer} \label{calculation}
\begin{figure}[t]
	\begin{center}
		\begin{subfigure}{0.45\textwidth}
			\centering
			\includegraphics[width=0.6\textwidth]{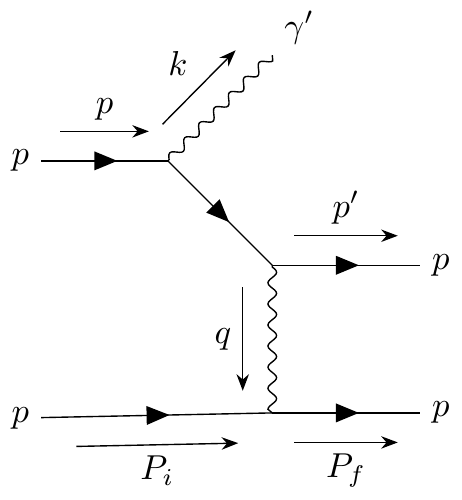}
			\caption{}
			\label{fig:diag1}
		\end{subfigure}
		\hfill
		\begin{subfigure}{0.45\textwidth}
			\centering
			\includegraphics[width=0.6\textwidth]{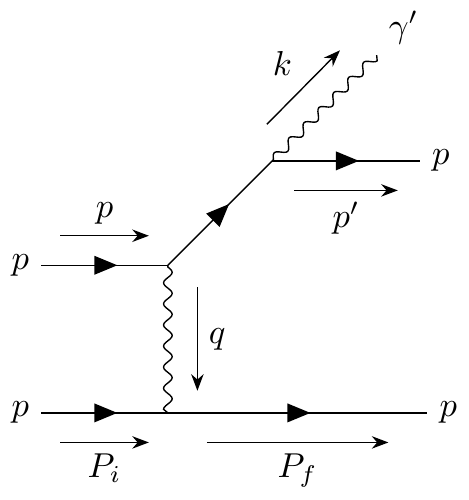}
			\caption{}
			\label{fig:diag2}
		\end{subfigure}
	\end{center}
	\caption{Feynman diagrams for proton bremsstrahlung.}
\end{figure}

In this section we calculate the matrix element of bremsstrahlung process according to the Feynman diagrams in figures \ref{fig:diag1}, \ref{fig:diag2}. Here we neglect the dark photon production by the target proton drawing an analogy with the case of photon emission in electron scattering considered in \cite{Altarelli:1964,Baier:1966jf,LL4}.
We extend the approach of ref.\,\cite{Blumlein:2013cua} to the region of the phase space with non-zero momentum transfer. 
For the initial particles, $P_i$ is the momentum of the target proton, $p$ is the momentum of the incident proton. The momenta of outgoing particles are $k$, $p'$ and $P_f$ for the dark photon, the outgoing and target protons respectively. The hypothetical virtual boson responsible for elastic pp scattering has the momentum $q$. In the lab frame the momenta can be written as 
\begin{align} \label{eq:momenta}
P_i^\mu &=\{M, 0, 0, 0\}, \\
p^\mu &=\{P+\frac{M^2}{2P}, 0, 0, P\}, \\
k^\mu &=\{zP+\frac{m^2_{\gamma '}+k^2_\perp}{2zP}, k_x, k_y, zP\}, \\
p'^\mu &=\{p'_0, -k_x-q_x, -k_y-q_y, P(1-z)-q_z\}, \\
q^\mu &=\{q_0, q_x, q_y, q_z\},
\end{align}
where
\begin{equation}
p'_0=P(1-z)+\frac{M^2+k^2_\perp}{2P\left(1-z\right)}+\frac{k_xq_x+k_yq_y}{P\left(1-z\right)}-q_z+\frac{q^2_\perp}{2P\left(1-z\right)}
\end{equation}
and the squares of transverse momenta are $k^2_\perp\equiv k_x^2+k_y^2$ and $q^2_\perp\equiv q_x^2+q_y^2$. Here we assume that $M/P \ll 1$ and 
\begin{align}
\frac{1}{P}\sqrt{\frac{k^2_\perp+m^2_{\gamma'}}{2}}& \ll z, \label{eq:zmin}\\
\frac{1}{P}\sqrt{\frac{k^2_\perp+M^2}{2}}& \ll 1-z. \label{eq:zmax}
\end{align}

While constructing the matrix element, for the hypothetical boson we take the photon propagator to connect currents of the incident ($L^\mu$) and the target ($J^\mu$) protons with charge $Q_b$ as follows
\begin{equation}
i{\cal M}=-i\epsilon e Q_b^2 L^\nu \frac{-i}{q^2} g_{\nu \lambda} J^\lambda\,. 
\end{equation}
Here the incident proton current reads
\begin{equation}
\label{L-current}
L_\nu= \epsilon^{*\mu}_{\gamma '}(k) \bar{u}(p') \left(\gamma_\nu \frac{\hat{p}-\hat{k}+M}{\left(p-k\right)^2-M^2}\gamma_\mu + \gamma_\mu\frac{\hat{k}+\hat{p'}+M}{\left(k+p'\right)^2-M^2}\gamma_\nu \right) u(p)
\end{equation}
and the target proton current forms the hadronic tensor $W^{\mu\nu}\equiv J^\mu J^{*\nu}$ that is parametrized in the standard way
by two form factors $W_1\left(\bar{s}, q^2 \right)$,
$W_2\left(\bar{s}, q^2\right)$ as  
\begin{equation} \label{eq:wmunu}
W^{\mu\nu}  = \left(-g^{\mu \nu}+\frac{q^\mu q^\nu}{q^2}\right)W_1 + \frac{1}{M^2}\left(P_i^\mu-\frac{q_\lambda P_i^\lambda}{q^2}q^\mu\right)\left(P_i^\nu-\frac{q_\rho P_i^\rho}{q^2}q^\nu\right)W_2.
\end{equation}
Applying the Dirac equation one can easily show that $q^\mu L_\mu = 0$, hence the absolute square of matrix element averaged over the initial states and summed over the final ones is
\begin{equation} \label{eq:Msquared}
\overline{|{\cal M}|^2} = \frac{1}{4}\epsilon^2e^2Q_b^4\frac{L_{\mu\nu}}{\left(q^2\right)^2}\left(-g^{\mu\nu}W_1+\frac{1}{M^2}P_i^\mu P_i^\nu W_2\right),
\end{equation}
where $L_{\mu \nu}\equiv L_\mu L^*_\nu$ is the hadronic tensor constructed of the incident proton current\,\eqref{L-current}.

Assuming that the functions $W_1$ and $W_2$ are of the same order of magnitude like in the case of structureless fermions, one can neglect the first term in \eqref{eq:Msquared} since in the lab frame 
\begin{equation}
\frac{1}{M^2}L_{\mu \nu} P_i^\mu P_i^\nu = L_{00}
\end{equation}
and, as it was shown in \cite{Kim:1973he}, that cancellations in \eqref{eq:Msquared} imply $g^{\mu\nu}L_{\mu\nu} \ll L_{00}$. Thus one only needs to explicitly calculate 
the 00 component of the incident proton hadronic tensor, which may be written as follows 
\begin{multline}
L^{\mu\nu}=-\frac{4}{S^2U^2}\left[g^{\mu\nu}\left(-SU\left(S^2+U^2\right)+q^2\left(2M^2(S+U)^2+m^2_{\gamma '}\left(S-U\right)^2+\right.\right.\right.
\\\left.\left.\left.+2SU\left(S+U\right)\right)-2q^4SU\right)-\right.
\\\left.-8M^2SUk^\mu k^\nu+4SU\left(m^2_{\gamma '}-S\right)p^\mu p^\nu+4SU\left(m^2_{\gamma '}-U\right)p'^\mu p'^\nu +\right.
\\\left.+2U\left(k^\mu p^\nu + p^\mu k^\nu \right)\left(2M^2\left(S+U\right)-m^2_{\gamma '}\left(S-U\right)+S^2-Sq^2\right)-\right.
\\\left.-2S\left(k^\mu p'^\nu + p'^\mu k^\nu \right)\left(2M^2\left(S+U\right)+m^2_{\gamma '}\left(S-U\right)+U^2-Uq^2\right)+\right.
\\\left.+2\left(p^\mu p'^\nu + p'^\mu p^\nu \right)\left(2M^2\left(S+U\right)^2+m^2_{\gamma '}\left(S^2+U^2\right)+SU\left(U+S\right)-2SUq^2\right)\right],
\end{multline}
with Lorentz invariants $U\equiv m^2_{\gamma '}-2p^\mu k_\mu$ and $S\equiv m^2_{\gamma '}+2p'^\mu k_{\mu}$. These invariants can be presented in terms of $H$ defined in \eqref{eq:H} as
\begin{align} 
U&=-H/z, \label{eq:U}\\
S&=\frac{H}{z(1-z)}+\frac{2}{1-z}\left(k_xq_x+k_yq_y\right)+\frac{q^2_\perp z}{1-z}, \label{eq:S}
\end{align}
where the latter equality is valid up to terms of the second order in $q$. 
 
Using the momentum conservation $p'=p-k-q$,  we extract in the numerator of $L^{00}$ terms up to the second order in $q$. 
Then we arrive at the approximation for $L^{00}$ (denoted by symbols with a hat) 
\begin{equation}
\label{L00}
\hat{L}^{00}=-\frac{4}{S^2U^2}\left(\hat{N}_{q^0}+\hat{N}_{q^1}+\hat{N}_{q^2}\right),
\end{equation}
\begin{equation}
\hat{N}_{q^0}\equiv \frac{H^4}{z^4\left(1-z\right)^3}\left(1+\left(1-z\right)^2\right),
\end{equation}
\begin{multline}
\hat{N}_{q^1}\equiv \frac{4q_0 H^3}{z^3\left(1-z\right)^2}P\left(1+\left(1-z\right)^2\right)+\left(k_xq_x+k_yq_y\right)\left[\frac{H^3}{z^3\left(1-z\right)^3}\left(6+2\left(1-z\right)^2\right)- \right.
\\\left. -\frac{8H}{z^2\left(1-z\right)}\left(2M^2+m^2_{\gamma '}\right)\left(m^2_{\gamma '}+k^2_\perp-z^2M^2\right)\right],
\end{multline}
\begin{multline}
\hat{N}_{q^2}\equiv \frac{4q^2H^2P^2}{z^2\left(1-z\right)}\left(1+\left(1-z\right)^2\right) + 16\left(k_xq_x+k_yq_y\right)^2\left(2M^2+m^2_{\gamma '}\right)P^2+
\\ 
+\frac{8q_0HP}{z\left(1-z\right)^2}\left(k_xq_x+k_yq_y\right)\left[\frac{H}{z}\left(2+\left(1-z\right)^2\right)-\left(2M^2+m^2_{\gamma '}\right)z\left(1-z\right)\right].
\end{multline}
Finally, we substitute \eqref{eq:U}, \eqref{eq:S} into prefactor of \eqref{L00} and obtain $\hat{L}^{00}\simeq -4\hat{B}$,
\begin{equation}
\hat{B}\equiv b_0+b_1q_0+b_2\left(k_xq_x+k_yq_y\right)+b_3\left(k_xq_x+k_yq_y\right)^2+b_4q^2+b_5q_0\left(k_xq_x+k_yq_y\right),
\end{equation}
where the coefficients $b_0,\dots,b_5$ take the form
\begin{align}
b_0&\equiv\frac{1+\left(1-z\right)^2}{1-z}, \label{eq:b0}\\
b_1&\equiv\frac{4zP}{H}\left(1+\left(1-z\right)^2\right), \\
b_2&\equiv-\frac{8\left(1-z\right)z^2}{H^3}\left(2M^2+m^2_{\gamma '}\right)\left(m^2_{\gamma '}+k^2_\perp-z^2M^2\right)-\\&\quad\, -\frac{2z\left(-1+\left(1-z\right)^2\right)}{H\left(1-z\right)},\\
b_3&\equiv\frac{16\left(1-z\right)^2 z^4}{H^4}\left(2M^2+m^2_{\gamma '}\right)P^2+\frac{z^2\left(-12+4\left(1-z\right)^2\right)}{H^2\left(1-z\right)}+\\&\quad\, +\frac{32}{H^4}\left(1-z\right)z^3\left(2M^2+m^2_{\gamma '}\right)\left(m^2_{\gamma '}+k^2_\perp-z^2M^2\right),  \label{eq:b3}\\
b_4&\equiv\frac{4\left(1-z\right)z^2P^2}{H^2}\left(1+\left(1-z\right)^2\right), \label{eq:b4}\\
b_5&\equiv-\frac{8z^2P}{H^2}\left(\left(1-z\right)^2+\frac{z^2}{H}\left(1-z\right)\left(2M^2+m^2_{\gamma'}\right)\right).
\end{align}

The differential cross section of the process $p(p)p(P_i)\rightarrow \gamma'(k)p(p')p(P_f)$ can be cast as 
\begin{equation} \label{eq:dsig}
\dd{\sigma}=\frac{\epsilon^2 \alpha Q_b^4}{32 MP \left(q^2\right)^2}\left(-\hat{B}\right)W_2\frac{\dd{k_\perp^2 \dd{z} \dd{\varphi_k}}}{\pi^2 z}\dd{\Phi_{p'P_f}},
\end{equation}
with the two-particle phase space volume 
\begin{equation} \label{eq:phasevol}
\dd{\Phi_{p' P_f}} = \frac{\dd[3]{p'}}{\left(2\pi\right)^3 2 p'_0} \frac{\dd[3]{P_f}}{\left(2\pi\right)^3 2 P_{f0}} \left(2\pi\right)^4 \delta^{(4)}\left(p+P_i-k-p'-P_f\right).
\end{equation}
One can rewrite \eqref{eq:dsig} by making use of the expression for $W_2$ derived in appendix\,\ref{elastic} as
\begin{equation}
\dd \sigma = \frac{\epsilon^2\alpha M \dd{k_\perp^2} \dd{z} \dd{\varphi_k}}{16\pi^2 zP \tilde{S}^2}\int \left(-\hat{B}\right)\abs{T_+(\bar{s}, q^2) + T_+^c(\bar{s}, q^2)}^2 \dd{\Phi_{p'P_f}},
\end{equation}
where $\sqrt{\bar{s}}$ is the center-of-mass energy of two protons exchanging the virtual hypothetical boson, $\tilde{S}^2\equiv\left(\bar{s}-2M^2\right)^2+q^2\left(\bar{s}-M^2\right)$, $T_+\left(\bar{s}, q^2\right)$ and $T_+^c\left(\bar{s}, q^2\right)$ are the fit functions for the differential cross section of elastic proton scattering \eqref{eq:elasticamp} defined as in \cite{ParticleDataGroup:2016lqr}.

The next step is to integrate over the phase space volume. Since in the lab frame $\vec{P_f}=\vec{q}$ we replace the second integration variable in \eqref{eq:phasevol} with $\dd^3 q$ and integrate over $\dd^3 p'$ using the $\delta$-function. Then we change the basis of coordinate system from the old one ($\vec{e}_x$, $\vec{e}_y$, $\vec{e}_z$) to the new basis ($\vec{n}_x$, $\vec{n}_y$, $\vec{n}_z$) of coordinate system with $z$ axis oriented along the vector $\vec{p}-\vec{k}$. The coordinates of vectors $\vec{n}_i$ in the basis $\vec{e}_j$ read
\begin{align}
\vec{n}_x &=N_x^{-1} \{P(1-z), 0, k_x\}, \\
\vec{n}_y &=N_y^{-1} \{-k_xk_y, k_x^2+P^2(1-z)^2, k_yP(1-z)\}, \\
\vec{n}_z &=N_z^{-1} \{-k_x, -k_y, P(1-z)\}, 
\end{align}
with  the normalization factors 
\begin{equation}
N_x^2\equiv P^2(1-z)^2+k_x^2, \hspace{5 mm} N_z^2\equiv P^2(1-z)^2+k_\perp^2, \hspace{5 mm} N_y^2\equiv N_x^2 N_z^2.
\end{equation}
Then integration over $\varphi_q$ gives
\begin{align}
\int\displaylimits_0^{2\pi}\left(k_xq_x+k_yq_y\right)\dd \varphi_q &= -2\pi N_z^{-1}|\vec{q}|k^2_{\perp}\cos \theta_q, \\
\int\displaylimits_0^{2\pi}\left(k_xq_x+k_yq_y\right)^2\dd \varphi_q &= \pi N_z^{-2}|\vec{q}|^2k_\perp^2 \left(P^2(1-z)^2\sin^2 \theta_q + 2k^2_\perp \cos^2 \theta_q\right).
\end{align}
The remaining $\delta$-function fixes the value of $\cos \theta_q$
\begin{equation}
\cos \hat{\theta}_q = \frac{1}{2|\vec{p}-\vec{k}||\vec{q}|}\left(\frac{H}{z}+\frac{t}{M}\left(P(1-z)+M+\frac{M^2}{2P}-\frac{m^2_{\gamma '}+k^2_{\perp}}{2zP}\right)\right),
\end{equation}
where $t\equiv -q^2$. Since the integrand does not depend on the azimuthal angle $\varphi_k$, we integrate over it and finally obtain
\begin{multline} \label{eq:bremcrsec}
\frac{\dd[2]{\sigma}}{\dd{k_\perp^2} \dd{z}} = \frac{\epsilon^2 \alpha}{32\left(2\pi \right)^2 z P \tilde{S}^2  \sqrt{P^2(1-z)^2+k_\perp^2}} \int\displaylimits_{t_\text{min}}^{t_\text{max}} \dd t \abs{T_+ + T_+^c}^2 \times \left[-b_0-\right. \\ \left. -\frac{b_1 t}{2M} + b_4 t+\left(b_2+\frac{b_5 t}{2 M}\right)\frac{k^2_\perp|\vec{q}|\cos \hat{\theta}_q}{\sqrt{P^2(1-z)^2 + k_\perp^2}}-\frac{b_3 k_\perp^2}{P^2(1-z)^2+k_\perp^2}\times\right.\\\left.\times\left(\frac{t}{2}\left(\frac{t}{4M^2}+1\right)P^2(1-z)^2+|\vec{q}|^2\cos^2\hat{\theta}_q\left(k_\perp^2 - \frac{P^2}{2}(1-z)^2\right)\right)\right],
\end{multline}
where the limits of integration are taken as in \cite{Kim:1973he}: $\sqrt{t_\text{max}}\sim m_{\gamma '}$ and 
\begin{equation}
\sqrt{t_\text{min}}=\frac{H}{2z(1-z)P}.
\end{equation}

In order to reproduce the cross section in the Weizsacker-Williams approximation one needs to keep in the integrand of \eqref{eq:bremcrsec} only terms proportional to $b_0$, $b_4$, the leading in $P^2$ part of $b_3$ multiplied by the leading and a subleading in $P^2$ part of the corresponding factor in \eqref{eq:bremcrsec},
\begin{equation}
I=\int\displaylimits_{t_\text{min}}^{t_\text{max}} \dd t \abs{T_+ + T_+^c}^2 \left(-b_0+b_4t-\frac{b_3}{2}k^2_\perp \left(t-t_{\text{min}}\right)\right)\,.
\end{equation}
Then, using \eqref{eq:b0}, \eqref{eq:b3} and \eqref{eq:b4} it is straightforward to show that 
\begin{equation} \label{eq:simI}
I\simeq \int\displaylimits_{t_\text{min}}^{t_\text{max}} \dd t \abs{T_+ + T_+^c}^2 A_{2\rightarrow 2} \frac{t-t_\text{min}}{2t_\text{min}},
\end{equation}
with $A_{2\rightarrow 2}$ defined in \eqref{eq:A22}. Thus if one omits relatively small terms dependent on the direction of transferred momentum, \eqref{eq:bremcrsec} will coincide with the cross section in the Weizsacker-Williams approximation \eqref{eq:Ritz} without form factors $F_\text{VMD}$ and $F_\text{virt}$.

\section{Numerical estimates and comparison} \label{comparison}
\begin{figure}[t]
	\begin{center}
		\begin{subfigure}{0.49\textwidth}
			\centering
			\includegraphics[width=1.03\textwidth]{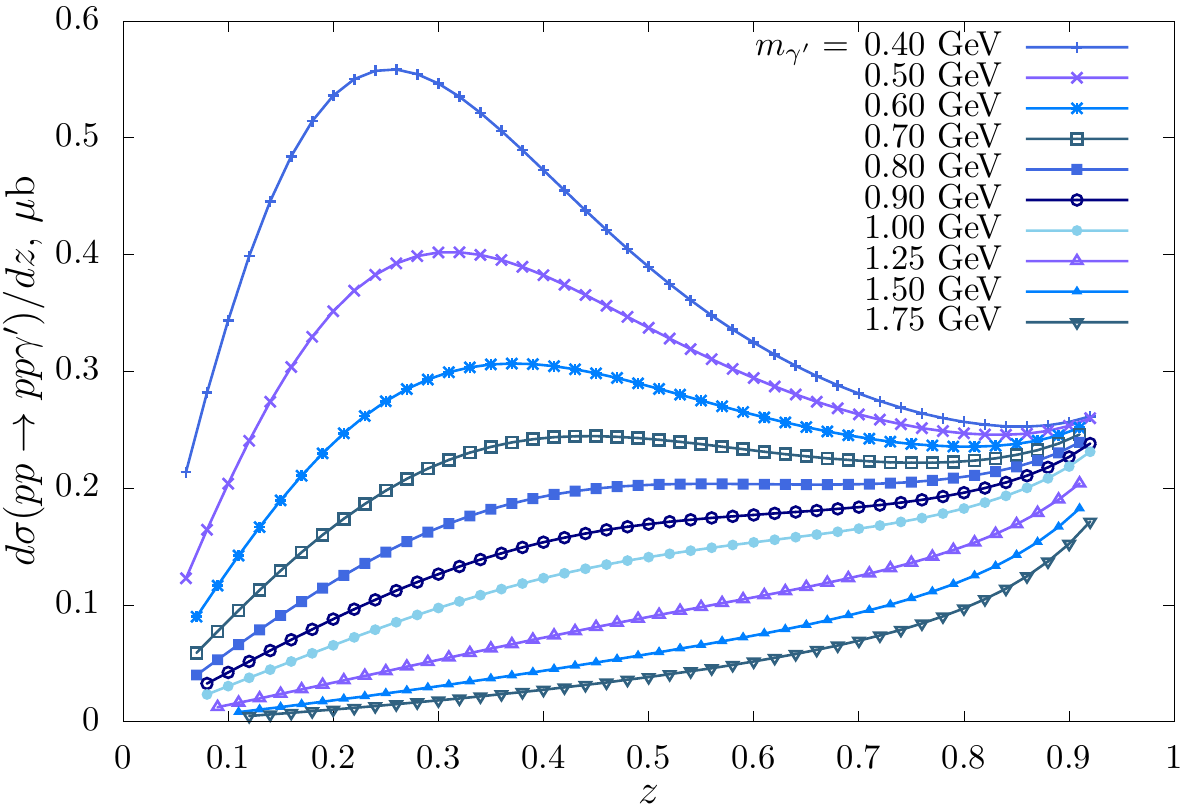}
			\caption{}
			\label{fig:brem-z}
		\end{subfigure}
		\hfill
		\begin{subfigure}{0.49\textwidth}
			\centering
			\includegraphics[width=1.03\textwidth]{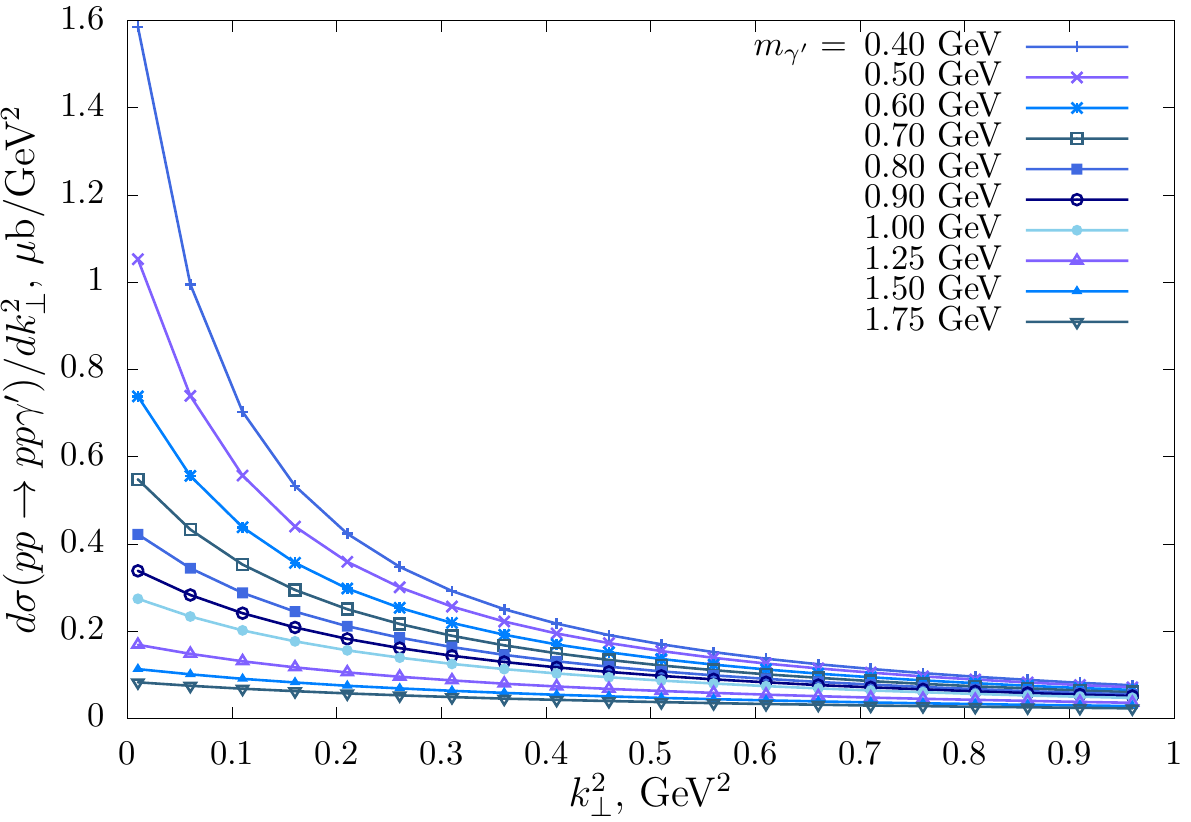}
			\caption{}
			\label{fig:brem-ktrsq}
		\end{subfigure}
	\end{center}
	\caption{The differential cross section of proton bremsstrahlung for different values of dark photon mass $m_{\gamma'}$ as a function of (a) ratio $z$ of the dark photon 3-momentum to the incident proton 3-momentum, $z\equiv k_z/p_z$, (b) the square of dark photon transverse momentum $k^2_\perp$. The incident proton momentum in the lab frame  $P=120\text{ GeV}$, dark photon masses are listed in the legend.}
\end{figure}

First, we present the numerical results obtained for the incident proton beam with the momentum in the lab frame $P=120\text{ GeV}$ as in the DUNE experiment under construction. The dependence of differential cross section \eqref{eq:bremcrsec} integrated over the momentum ratio $z$ or over the square of dark photon momentum $k^2_{\perp}$ on dark photon mass is shown in figures\,\ref{fig:brem-z}, \ref{fig:brem-ktrsq}.
We integrate over $k^2_\perp$ in the region 0.01-1 GeV$^2$ and over $z$ in the region obeying the assumptions \eqref{eq:zmin}, \eqref{eq:zmax}. One can observe that the differential cross section in figure~\ref{fig:brem-z} exhibits a maximum in the $z$ range 0.2-0.4 for dark photons lighter than $0.8$\,GeV. For heavier dark photons the behavior of the cross section changes and the maximum is reached at the maximal $z$. Figure \ref{fig:brem-ktrsq} shows the rapid decrease in the differential cross section with growing transverse dark photon momentum $k_\perp$. Note that the dark photon distribution over the transverse momentum is important, since it defines the outgoing trajectories which show which part of the produced dark photons can reach the downstream detector, that is a typical setup adopted e.g. in SHiP, DUNE and other experiments planning to search for the dark photon.

\begin{figure}[t]
	\begin{center}
		\begin{subfigure}{0.49\textwidth}
			\centering
			\includegraphics[width=1.03\textwidth]{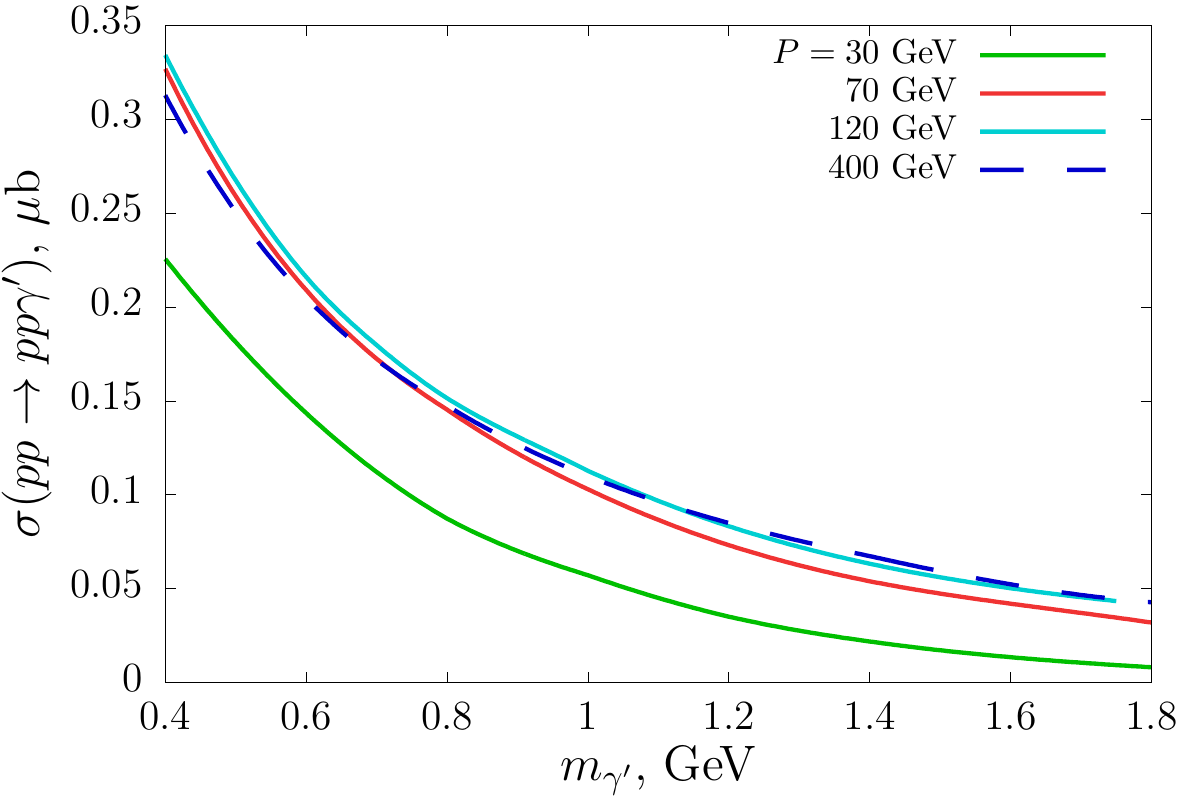}
			\caption{}
			\label{fig:full-crsec-P}
		\end{subfigure}
		\hfill
		\begin{subfigure}{0.49\textwidth}
			\centering
			\includegraphics[width=1.03\textwidth]{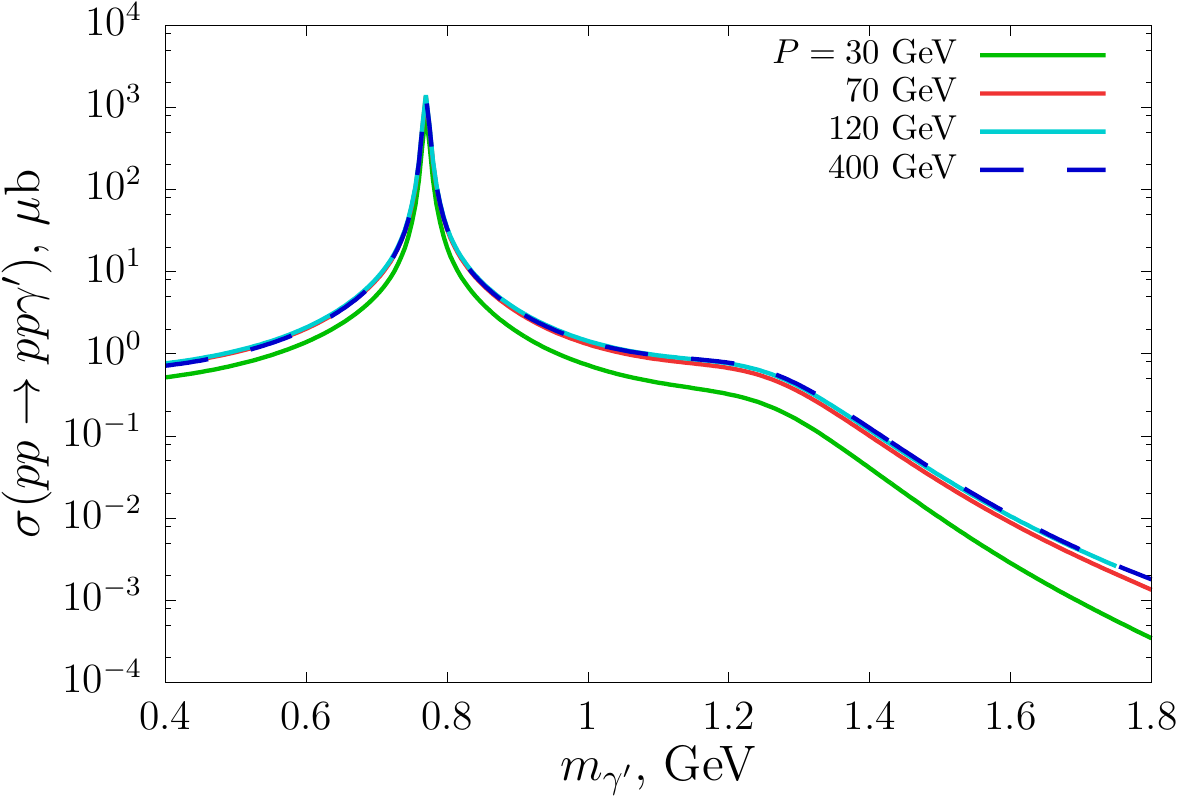}
			\caption{}
			\label{fig:full-crsec-P-VMD}
		\end{subfigure}
	\end{center}
	\caption{Full proton bremsstrahlung cross section as a function of dark photon mass for the values of incident proton momenta in the lab frame $P$ at J-PARC (green line), U-70 (red line), Fermilab (light blue line), CERN SPS (dark blue line). In figure~\ref{fig:full-crsec-P} the cross section is shown \emph{without} a non-trivial contribution of VMD form factor, i.e. we set $F_\text{VMD}=1$. The pronounced peak in figure~\ref{fig:full-crsec-P-VMD} at $m_{\gamma'}\approx0.8$\,GeV and the knee at $m_{\gamma'}\approx1.3$\,GeV are induced by the form factor $F_{\text{VMD}}$ \eqref{eq:vmdff} accounting for the dark photon mixing with vector meson states, see table~\ref{tab:VMD}.}
\end{figure}

Second, the dependence of the full integrated bremsstrahlung cross section on dark photon mass for several phenomenologically interesting values of $P$ is outlined in figures\,\ref{fig:full-crsec-P},\,\ref{fig:full-crsec-P-VMD}. 
Figure~\ref{fig:full-crsec-P} shows the differential cross section \eqref{eq:bremcrsec} integrated over $k_\perp^2$ and $z$. In figure~\ref{fig:full-crsec-P-VMD} we further multiply the result by the absolute square of VMD form factor \eqref{eq:vmdff}. The bremsstrahlung cross section for proton beam with $P=30$\,GeV in the lab frame produced by the Main Ring of J-PARC facility is shown in green. Red line demonstrates the bremsstrahlung cross section of the proton beam with $P=70$\,GeV at proton synchrotron U-70. The full cross section for Fermilab proton beam with $P=120$\,GeV is shown in light blue. Dark blue line corresponds to the CERN SPS proton beam with $P=400$\,GeV. The differential cross section \eqref{eq:bremcrsec} feebly depends on $P$, since the $1/P^2$ integral prefactor behaviour gets compensated by the integral $I$ proportional to $P^2$ which is evident from \eqref{eq:simI}. Thus the dependence of the cross section on $P$ is mostly due to varying limits of integration \eqref{eq:zmin}, \eqref{eq:zmax}. This dependence vanishes at large $P$, as is demonstrated by light and dark blue lines in figure~\ref{fig:full-crsec-P}.

\begin{figure}[t]
	\begin{center}
		\begin{subfigure}{0.48\textwidth}
			\centering
			\includegraphics[width=1.03\textwidth]{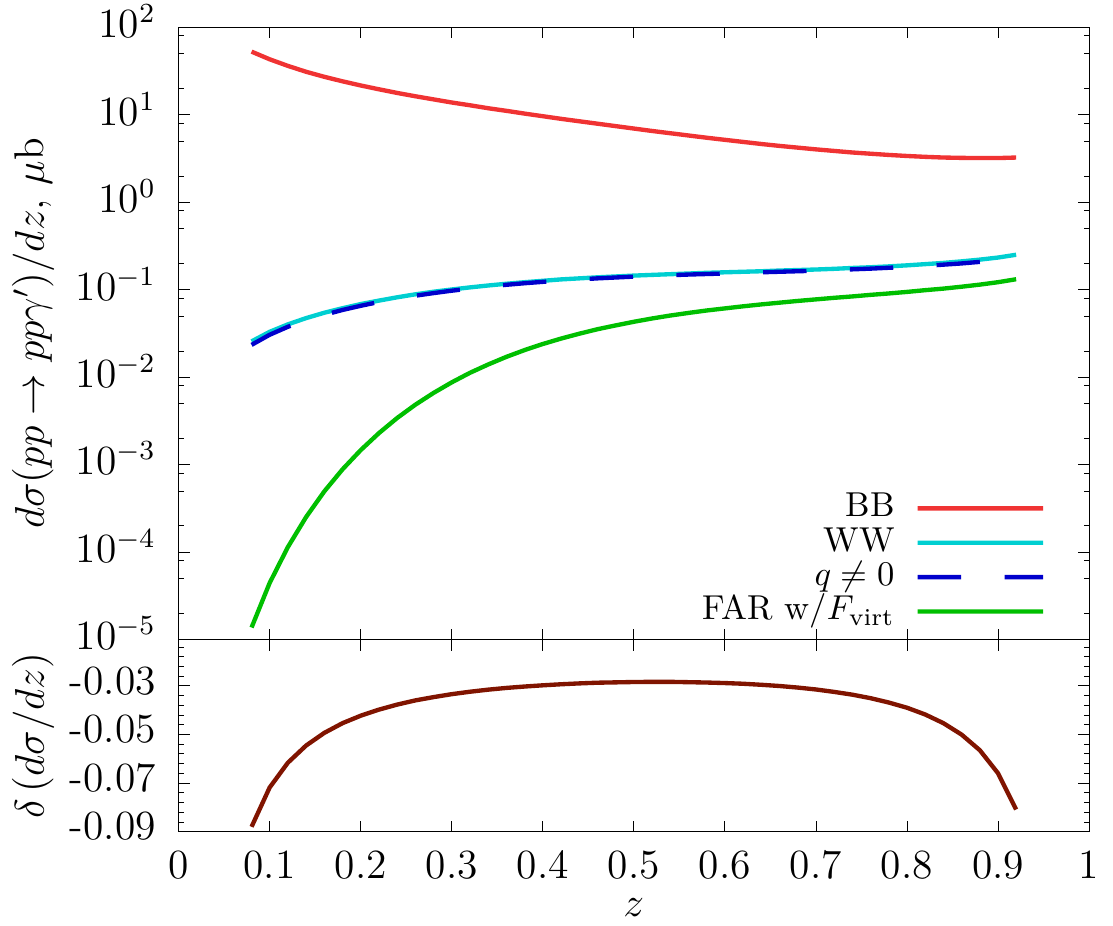}
			\caption{}
			\label{fig:brem-z-comp}
		\end{subfigure}
		\hfill
		\begin{subfigure}{0.5\textwidth}
			\centering
			\includegraphics[width=1.03\textwidth]{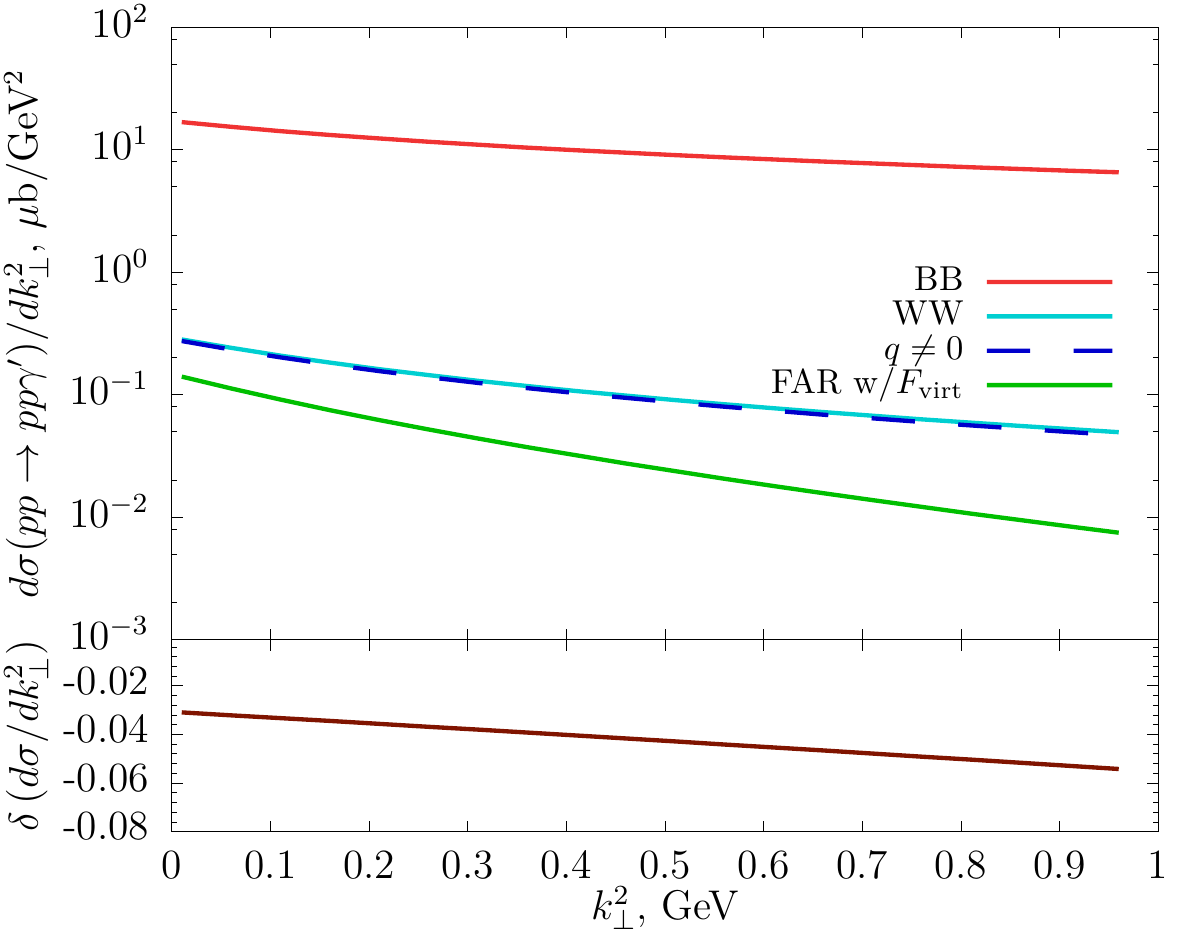}
			\caption{}
			\label{fig:brem-ktrsq-comp}
		\end{subfigure}
	\end{center}
	\caption{At the top: the differential cross section of proton bremsstrahlung according to the result of this work (dark blue line), Blumlein and Brunner (red line) \cite{Blumlein:2013cua}, Foroughi-Abari and Ritz (green line, includes the off-shell hadronic form factor $F_\text{virt}$ \eqref{eq:virtff}) \cite{Foroughi-Abari:2021zbm} and in the Weizsacker-Williams approximation (light blue line) \cite{Kim:1973he}; at the bottom: the relative deviation of the differential cross section obtained in this work from the WW approximation as a function of (a) the ratio $z$ of the dark photon momentum to the incident proton momentum, (b) the square of dark photon transverse momentum $k^2_\perp$. The incident proton momentum in the lab frame  $P=120\text{ GeV}$, dark photon mass $m_{\gamma'}=1\text{ GeV}$.}
\end{figure}

In figures~\ref{fig:brem-z-comp},\,\ref{fig:brem-ktrsq-comp}
we fix $m_{\gamma'}=1\text{ GeV}$ and compare the bremsstrahlung differential cross sections: the result of this work, \eqref{eq:bremcrsec} integrated over $z$ or $k^2_\perp$ (dark blue line), agrees well (within 3--9\%) with the Weizsacker-Williams (WW) approximation that is shown in light blue and is given by \eqref{eq:Ritz} with unity form factors $F_\text{VMD}$ and $F_\text{virt}$ \cite{Kim:1973he}. The red line presents the answer \eqref{eq:BB1}-\eqref{eq:BB2} of Blumlein and Brunner (BB) \cite{Blumlein:2013cua}, where the splitting function \eqref{eq:BB2} is multiplied by the \emph{elastic} $pp$-cross section \eqref{eq:elasticamp} integrated over $q^2$. The result of Foroughi-Abari and Ritz (FAR) \cite{Foroughi-Abari:2021zbm}, where we include in \eqref{eq:Ritz} the off-shell hadronic form factor $F_\text{virt}$ \eqref{eq:virtff} and still put $F_\text{VMD}=1$, is shown in green. If one puts $F_\text{virt}=1$, the result of FAR \eqref{eq:Ritz} coincides with the WW approximation. While the result of BB exceeds our result in figure~\ref{fig:brem-z-comp} by 1-3 orders of magnitude, the result of FAR is smaller than the latter. Since from \eqref{eq:simI} it follows that our result should agree well with the WW approximation, in the bottom of figures~\ref{fig:brem-z-comp},\,\ref{fig:brem-ktrsq-comp} we plot the relative deviation of the differential cross section, obtained in this work from the WW answer
\begin{equation}
    \delta\left(\frac{\dd\sigma}{\dd z}\right)\equiv \left. \left(\frac{\dd\sigma}{\dd z}-\left(\frac{\dd\sigma}{\dd z}\right)_\text{WW}\right)\middle/{\left(\frac{\dd\sigma}{\dd z}\right)_\text{WW}} \right.,
\end{equation}
and the analogous one for the differential cross section dependent on $k^2_\perp$, $\delta\left({\dd\sigma}/{\dd k^2_\perp}\right)$.

It is important to note that these three answers show completely different behavior at small $z$: the growth of BB cross section is due to the term proportional to $1/z$ in \eqref{eq:BB2}, and the off-shell hadronic form factor $F_\text{virt}$, see eq.\,\eqref{eq:virtff} suppresses the FAR cross section in the limit $z\rightarrow 0$. Although this discrepancy still needs to be resolved, the $1/z$ singularity in the BB cross section looks very unnatural in comparison with WW approximation and our result which is obtained considering the diagram of $2\rightarrow3$ bremsstrahlung process as a whole, without any splitting into auxiliary subprocesses.  

The differential cross section obtained in this work agrees not only with the WW approximation, but also with the result of FAR, if one does not multiply the latter by the absolute square of the off-shell hadronic form factor $F_\text{virt}$ \eqref{eq:virtff}. We should also note that Foroughi-Abari and Ritz used the pomeron propagator with the non-integer power of $q^2$ and the proton-pomeron couplings obtained within the framework of Donnachie-Landshoff pomeron model. In contrast, we utilize the photon-type propagator and derive the value of hypothetical boson-proton-proton coupling from the experimental fit for differential cross section of elastic $pp$-scattering. Despite that our approaches are, obviously, different, the results obtained in the WW approximation, our work and by FAR coincide numerically within about 10\%. In addition, Foroughi-Ababi and Ritz performed \cite{Foroughi-Abari:2021zbm} another calculation of the elastic bremsstrahlung with the one-pomeron exchange, that also exhibits a close agreement with the WW approximation.

\begin{figure}[t]
	\begin{center}
		\begin{subfigure}{0.49\textwidth}
			\centering
                \includegraphics[width=1.03\textwidth]{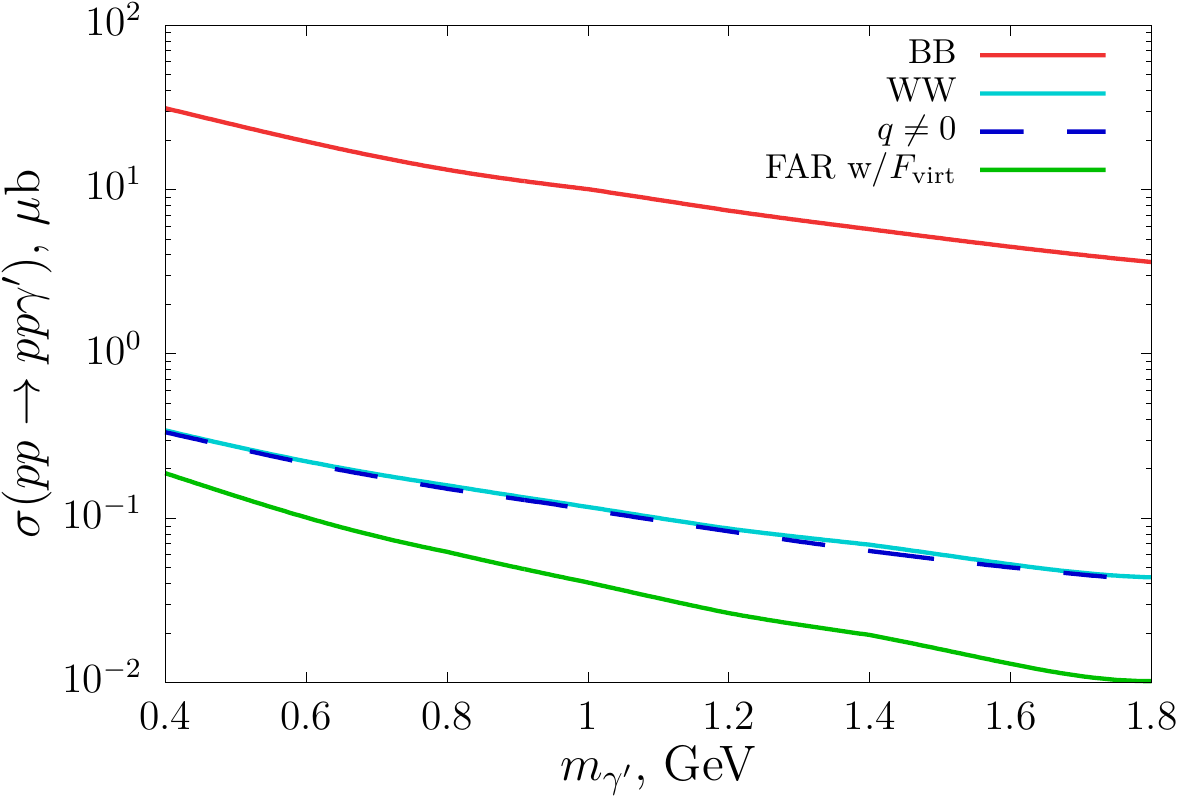}
			\caption{}
			\label{fig:full-crsec}
		\end{subfigure}
		\hfill
		\begin{subfigure}{0.49\textwidth}
			\centering
			\includegraphics[width=1.03\textwidth]{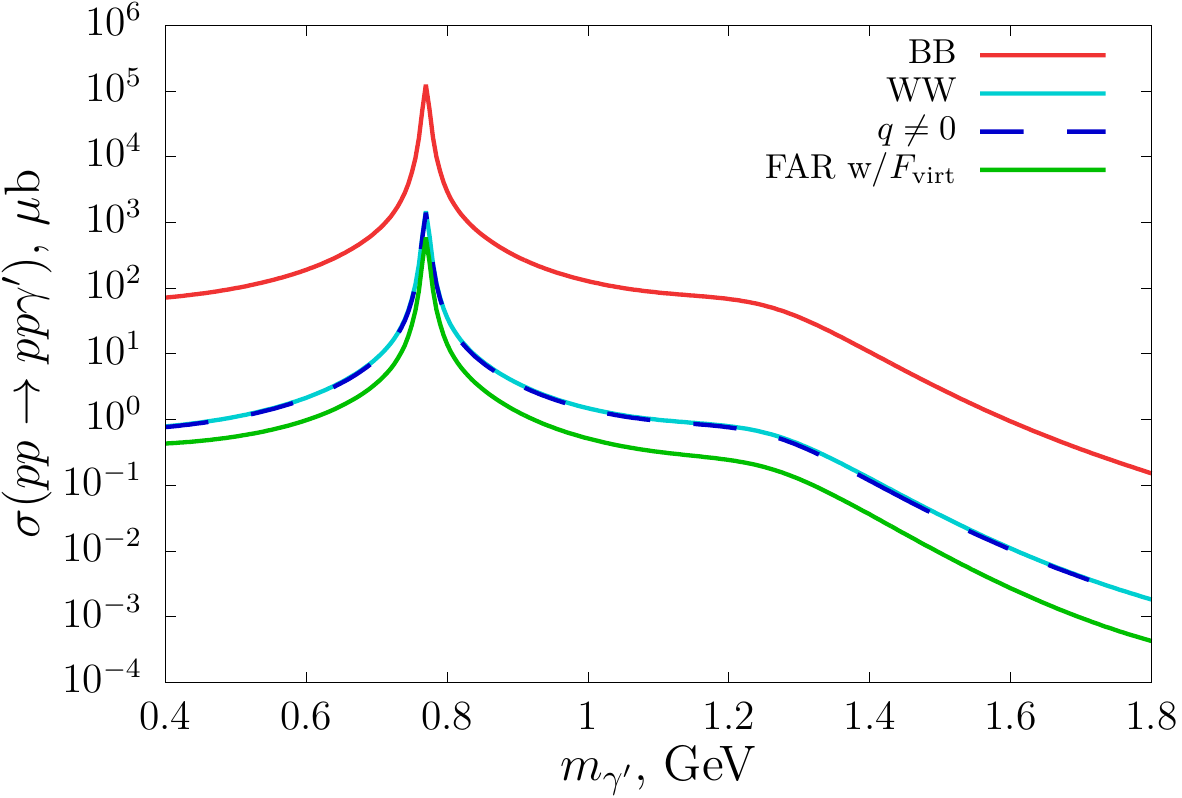}
			\caption{}
			\label{fig:full-crsec-VMD}
		\end{subfigure}
	\end{center}
	\caption{Full proton bremsstrahlung cross sections as functions of dark photon mass (a) \emph{without} the VMD form factor, (b) with the VMD form factor $F_\text{VMD}$ \eqref{eq:vmdff}. The color scheme is the same as in figures~\ref{fig:brem-z-comp},\,\ref{fig:brem-ktrsq-comp}. The incident proton momentum in the lab frame is $P=120\text{ GeV}$.}
\end{figure}

Figures~\ref{fig:full-crsec},\,\ref{fig:full-crsec-VMD}
show the full integrated bremsstrahlung cross sections as functions of the dark photon mass. In figure~\ref{fig:full-crsec} we omit the VMD form factor putting $F_\text{VMD}=1$, while in figure~\ref{fig:full-crsec-VMD} we multiply all the results by the absolute square of the same form factor $F_\text{VMD}$ \eqref{eq:vmdff}. The result of this work is shown in dark blue, the red line represents the answer of BB, the answer of FAR is indicated by the green line. Our result with non-zero momentum transfer agrees very well with the bremsstrahlung cross section in the WW approximation (light blue line) and deviates from the FAR answer by a factor 2-5, while the BB computation greatly exceeds both of them.

We also show in the appendix~\ref{contour} how the sensitivity of the SHiP experiment ($P=400\text{~GeV}$) depends on the approach chosen for estimating the dark photon production cross section. There we revisit the contours of the potentially excluded regions obtained earlier in \cite{Gorbunov:2014wqa} and \cite{SHiP:2020vbd}. Following these works we consider only the proton bremsstrahlung and assume nearly null background conditions provided by the SHiP muon shield. Our results are summarized in figure~\ref{fig:contour}. For the given dark photon mass $m_{\gamma^\prime}$ the excluded value of kinetic mixing $\epsilon$ can vary up to 1 order of magnitude. We stress that this is not the final answer for the sensitivity of the SHiP experiment, since following the original work of SHiP collaboration \cite{SHiP:2020vbd} we did not combine these results with the production via meson decays and Drell-Yan process. It will be also important to consider the inelastic bremsstrahlung (for further discussion see section~\ref{discussion}). We would like to draw the attention of the researchers planning searches for dark photons at the fixed-target experiments to this issue: the applicability of BB approximation should be first studied thoroughly.
\section{Discussion} \label{discussion}
To summarize, we have obtained the elastic bremsstrahlung cross section considering the non-zero momentum transfer between incident and target protons. Our result agrees well with the Weizsacker-Williams approximation. That is the Weizsacker-Williams approximation is applicable not only to structureless fermions, but also to protons and can give reliable results. With our method we present the refined predictions for the dark photon production cross section to be used for the data analysis with proton beams of 30\,GeV, 70\,GeV, 120\,GeV and 400\,GeV. 

Following the well-known study of the proton bremsstrahlung \cite{Blumlein:2013cua}, in this paper we considered \emph{elastic} proton scattering as the protons exchanging a hypothetical boson coupled to them like a massless vector, with the photon-type propagator. Generally speaking, the elastic proton scattering should be described as the exchange of effective particle, pomeron, with a completely different propagator dependence on 4-momentum squared $q^2$ (see \cite{Donnachie:2002en} for a review). The pomeron model, closest to our rough consideration, is the Donnachie-Landshoff (DL) model \cite{Donnachie:1983hf,Donnachie:1985iz,Donnachie:1987gu}. The DL pomeron is a vector with the charge parity $C=+1$. For a long time the DL model was successfully used to describe the elastic proton scatterings as it was one of the first proposed models, being simple, easy to use and adequately describing the experimental data available at the time  \cite{Donnachie:2002en}. Later it has been criticised for poor grounding in QFT and has been argued to be ruled out by experiment \cite{Britzger:2019lvc}. However, there are alternative pomeron models considering it as an effective spin-0 or spin-2 particle \cite{Ewerz:2013kda}. The scalar nature of pomeron is argued to be excluded by the STAR experiment and tensor pomeron model is suggested to be the most suitable \cite{Ewerz:2016onn}. Recent works \cite{Lebiedowicz:2022nnn,Lebiedowicz:2023mhe} contain the calculations of the SM bremsstrahlung and the central-exclusive production cross sections within the tensor pomeron model. One of the crucial future tasks is to revisit the dark photon production cross section in the framework of contemporary pomeron models.

In order to estimate the dark photon production cross section in the bremsstrahlung taking place in a real fixed-target experiment, one needs to consider the process $pA\rightarrow pA\gamma'$, where $A$ is the target nuclei. Following \cite{Watanabe:2023rgp}, we assume that the hypothetical vector particle considered in our study has equal coupling constants with proton and neutron ($Q_b$ charge in our calculation) just like a pomeron. Then the roughest estimate for the cross section of proton bremsstrahlung on a target composed of atoms with the number of nucleons $A$ is $\sigma\left(pA\rightarrow pA\gamma'\right)=A\times\sigma\left(pp\rightarrow pp\gamma'\right)$.

The cross section of the inelastic bremsstrahlung is expected to be an order of magnitude greater than the elastic bremsstrahlung cross section that we studied in this paper. It is the inelastic process that mostly contributes to dark photon production. Therefore it is also important to explore the inelastic dark photon production. From general considerations there should be the possibility to break the bremsstrahlung cross section into the proton (in)elastic scattering cross section and the splitting probability as it has been done by Blumlein and Brunner in \cite{Blumlein:2013cua}. But then it is not clear why the result for elastic cross section obtained in this splitting approach is so different from the Weizsacker-Williams approximation and our non-zero momentum transfer calculation. One of the questions for future research is whether the inelastic bremsstrahlung cross section calculated taking into account non-zero momentum transfer indeed rapidly grows at small $z$ as is stated in \cite{Blumlein:2013cua} or not. The task to resolve the differences between existing approaches requires further consideration.

\acknowledgments
We are indebted to V.\,Kim, I.\,Timiryasov, O.\,Teryaev and V.\,Vechernin for valuable discussion and correspondence. 
The work is partially supported by the  
Russian Science Foundation RSF grant 21-12-00379. The work of EK is supported by the grant of ``BASIS" Foundation no. 21-2-10-37-1.

\appendix
\section{Elastic pp scattering} \label{elastic}
In this appendix we consider the elastic proton scattering $p\left(p'+q\right)p\left(P_i\right)\rightarrow p\left(p'\right)p\left(P_f\right)$.  The fit of differential cross section to experimental data is usually parameterized as follows
\begin{equation} \label{eq:elasticamp}
\dv{\sigma_{pp}}{q^2}=\frac{\abs{T_+(\bar{s}, q^2)+T_+^c(\bar{s}, q^2)}^2}{16\pi \bar{s}\left(\bar{s}-4M^2\right)},
\end{equation}
where the Mandelstam variable is $\overline{s}\equiv \left(p'+q+P_i\right)^2$, functions $T_+(\bar{s}, q^2)$ and $T_+^c(\bar{s}, q^2)$ are defined as in \cite{ParticleDataGroup:2016lqr} and the numerator is the absolute square of matrix element averaged over initial and summed over final spins, 
\begin{equation}\label{eq:M2viaT-T}
 \overline{\abs{\mathcal{M}}^2}(\bar{s}, q^2)\equiv  \abs{T_+(\bar{s}, q^2)+T_+^c(\bar{s}, q^2)}^2.  
\end{equation}
The matrix element constructed along the same lines as for the proton bremsstrahlung reads
\begin{equation}
i\mathcal{M}=\frac{iQ_b^2}{q^2}\bar{u}\left(p'\right)\gamma_\mu u\left(p'+q\right)J^\mu,
\end{equation}
where $J^\mu$ is the target proton current. Then the averaged matrix element square is 
\begin{equation} \label{eq:matrelemsq}
\overline{\abs{\mathcal{M}}^2}=\frac{1}{4}\frac{Q_b^4}{\left(q^2\right)^2}\widetilde{L}_{\mu \nu}W^{\mu \nu},
\end{equation}
with the hadronic tensor of incident proton 
\begin{equation}
\widetilde{L}_{\mu \nu} = 4\left(2p'_\mu p'_\nu + p'_\mu q_\nu + p'_\nu q_\mu - g_{\mu \nu} p'^\lambda q_\lambda \right)
\end{equation}
and target proton hadronic tensor $W^{\mu \nu}$ defined in the main text by eq.\,\eqref{eq:wmunu}. Since the incident proton hadronic tensor is transversal, $q^\mu \widetilde{L}_{\mu \nu}=0$, for the product of hadronic tensors one obtains
\begin{equation} \label{eq:lmunuwmunu2}
\widetilde{L}_{\mu\nu}W^{\mu\nu}=-g^{\mu\nu}\widetilde{L}_{\mu\nu}W_1 + \frac{1}{M^2}P_i^\mu P_i^\nu \widetilde{L}_{\mu \nu} W_2.
\end{equation}
Let us assume that the functions $W_1$ and $W_2$ are of the same order of magnitude as in the case of structureless fermions. Then for typical values of $M$, $q^2$ and $\bar s$ the first term in \eqref{eq:lmunuwmunu2} is numerically much less than the second one, and hence can be neglected, which yields from eqs.\,\eqref{eq:M2viaT-T} and \eqref{eq:matrelemsq} 
\begin{equation}
W_2\left(\bar{s}, q^2\right)=\frac{2\left(q^2\right)^2M^2}{Q_b^4}\frac{\abs{T_+(\bar{s}, q^2) + T_+^c(\bar{s}, q^2)}^2}{\left(\bar{s}-2M^2\right)^2+q^2\left(\bar{s}-M^2\right)}\,.
\end{equation} 

\section{The sensitivity of SHiP experiment to dark photons} \label{contour}

Due to the kinetic mixing with the SM photons in the Lagrangian \eqref{eq:Lagr}, the dark photon can decay into pairs of charged leptons with the decay width \cite{Miller:2021ycl}
\begin{equation} \label{eq:decll}
    \Gamma(\gamma^\prime\rightarrow l^+ l^-) = \frac{\alpha \epsilon^2}{3} m_{\gamma^\prime} \sqrt{1-\frac{4m_l^2}{m_{\gamma^\prime}^2}}\left(1+\frac{2m_l^2}{m^2_{\gamma^\prime}}\right),
\end{equation}
where $m_l$ is the lepton mass. The decay width to hadrons, 
\begin{equation}
    \Gamma(\gamma^\prime\rightarrow \text{hadrons}) = \frac{\alpha \epsilon^2}{3} m_{\gamma^\prime} R(m_{\gamma^\prime}) \sqrt{1-\frac{4m_\mu^2}{m_{\gamma^\prime}^2}}\left(1+\frac{2m_\mu^2}{m^2_{\gamma^\prime}}\right)\,,
\end{equation}
explicitly depends on the experimentally-determined R-ratio \cite{ParticleDataGroup:2022pth}
\begin{equation}
    R(\sqrt{s})\equiv\frac{\sigma(e^+e^-\rightarrow \text{hadrons})}{\sigma(e^+e^-\rightarrow \mu^+\mu^-)}.
\end{equation}
In this appendix we consider the dark photon which can decay only to visible SM final states while its decay into invisible final states with dark matter is, e.g., forbidden by the kinematics. Then the total dark photon decay width reads
\begin{equation}
    \Gamma_\text{tot}= \Gamma(\gamma^\prime\rightarrow e^+ e^-) + \Gamma(\gamma^\prime\rightarrow \mu^+ \mu^-) + \Gamma(\gamma^\prime\rightarrow \text{hadrons}).
\end{equation}

The decay length for the flux of dark photons moving with the relativistic factor $\gamma$ and velocity $\beta\equiv k_z/k_0$ is
\begin{equation}
    L(z, k^2_\perp) = \frac{\gamma \beta}{\Gamma_\text{tot}}. 
\end{equation}
So the probability for a dark photon to decay inside the fiducial volume of SHiP detector is
\begin{equation}
    w_\text{det}(z, k^2_\perp) = e^{-l_\text{sh}/L(z, k^2_\perp)} \left(1-e^{-l_\text{det}/L(z, k^2_\perp)}\right),
\end{equation}
where we took the conservative estimates from \cite{Gorbunov:2014wqa} for shielding length $l_\text{sh}=60\text{~m}$ and the length of detector fiducial volume $l_\text{det}=50\text{~m}$. It allows us to illustrate the impact of the cross section on the estimate of the SHiP sensitivity to the model parameters. Namely, we compare the sensitivity of SHiP experiment based on our result of the hidden photon production with the estimates previously obtained in \cite{Gorbunov:2014wqa} using the BB answer.
Following \cite{Gorbunov:2014wqa} we select only dark photons which have the momenta lying inside the cone with the angle
\begin{equation}
    \theta_\text{cr} = \frac{r}{l_\text{sh}+l_\text{det}},
\end{equation}
where the cone radius is $r=2.5\text{~m}$.

Finally, we take the estimate for the luminosity of the SHiP experiment from \cite{SHiP:2020vbd}
\begin{equation} \label{eq:lum}
    \mathcal{L}_\text{SHiP}=\frac{N_\text{POT}}{\sigma^\text{inel}_\text{SHiP}}
\end{equation}
with $N_\text{POT}=10^{20}$ protons on target and the inelastic proton-nucleon cross section $\sigma^\text{inel}_\text{SHiP}=10.7\text{~mb}$. 

Combining the expressions \eqref{eq:decll}--\eqref{eq:lum} gives the total number of signal events
\begin{equation}
    N_{\gamma^\prime}=\mathcal{L}_\text{SHiP}|F_\text{VMD}|^2 \int \dd k^2_\perp \dd z \frac{\dd^2 \sigma}{\dd k^2_\perp \dd z} w_\text{det}(z, k^2_\perp) H\left(\theta_\text{cr}-\frac{k_\perp}{zP}\right),
\end{equation}
where $H(x)$ is the Heaviside step-function equal to 1 for $x>0$ and to 0 for $x<0$. 

\begin{figure}[t]
	\begin{center}
		\includegraphics[width=0.6\textwidth]{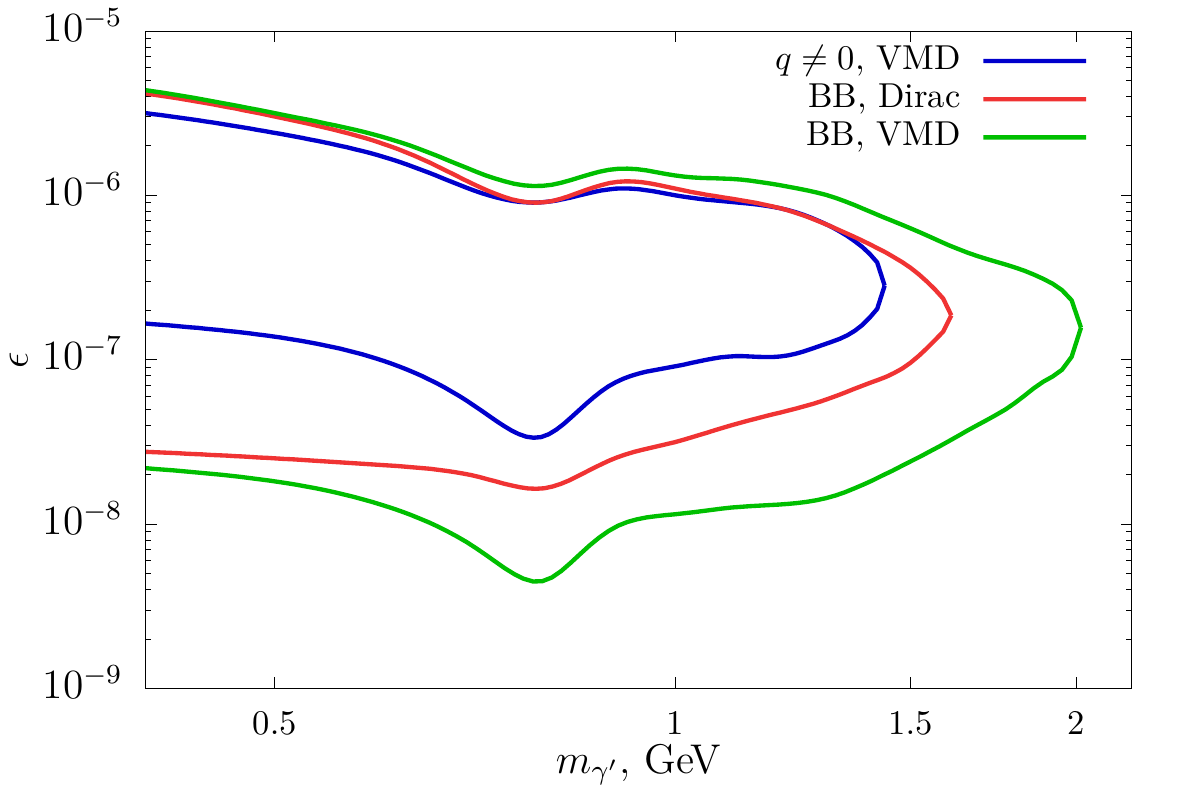}
		\caption{Contours of the regions in the dark photon parameter space expected to be explored by the SHiP experiment for the elastic cross section obtained in this work with the VMD form factor $F_\text{VMD}$ \eqref{eq:vmdff} (blue), the BB inelastic cross section \eqref{eq:BB1} with the Dirac form factor $F_D$ \eqref{eq:diracff} (red) and the VMD form factor $F_\text{VMD}$ \eqref{eq:vmdff} (green).}
		\label{fig:contour}
	\end{center}
\end{figure}

Figure~\ref{fig:contour}
shows the contours in the parameter space $(m_{\gamma^\prime}, \epsilon)$ corresponding to 3 signal events inside the SHiP detector. The blue curve corresponds to the number of events estimated using the elastic bremsstrahlung cross section obtained in this work \eqref{eq:bremcrsec} and VMD form factor $F_\text{VMD}$ \eqref{eq:vmdff}. To compare with the results of  \cite{SHiP:2020vbd}, we also show the contour of potentially excluded regions evaluated with the BB cross section \eqref{eq:BB1} and the VMD form factor $F_\text{VMD}$ \eqref{eq:vmdff} (green line). The conservative expected exclusion limits analogous to the obtained in \cite{Gorbunov:2014wqa} are shown in red and were computed using the BB cross section \eqref{eq:BB1} and the Dirac form factor \cite{SHiP:2020vbd}
\begin{equation} \label{eq:diracff}
    F_D(m_{\gamma^\prime})= 
    \begin{cases}
        \frac{\Lambda^4}{m_{\gamma^\prime}^4}, & m_{\gamma^\prime} > \Lambda,\\
        1, & m_{\gamma^\prime} < \Lambda,
    \end{cases}
\end{equation}
where $\Lambda^2=0.71~\text{GeV}^2$. The non-observation of signal events for parameter values inside the contours means the exclusion of these regions at the 95\% CL. The proposed construction of muon shield in the SHiP experiment allows one to neglect the possible background \cite{SHiP:2020vbd}. One can see that changes in the predicted experimental sensitivity, depending on the model of dark photon production, are quite serious, that  warns the experimental community.

\end{document}